\documentclass[prb,twocolumn,floatfix,superscriptaddress,showpacs]{revtex4}
\usepackage{amsfonts}
\usepackage{amsmath}
\usepackage{bm}
\usepackage{graphicx}
\begin{document}
\title{Infrared Hall angle in the $d$-density wave state: a comparison of theory and experiment}
\author{ Sumanta Tewari}
\affiliation
{Department  of Physics and Institute of Theoretical Science, University of Oregon, Eugene,
Oregon 97403--5203, USA }
\author{ Sudip Chakravarty}
\affiliation
{Department of Physics and Astronomy, University of California at Los Angeles,
Los Angeles, California 90095-1547, USA}
\author{ John Ove Fj{\ae}restad}
\affiliation
{Department of Physics and Astronomy, University of California at Los Angeles, Los Angeles,
California 90095-1547, USA}
\author{Chetan Nayak}
\affiliation
{Department of Physics and Astronomy, University of California at Los Angeles,
Los Angeles, California 90095-1547, USA}
\author{ Richard S. Thompson}
\affiliation
{Department of Physics and Astronomy, University of Southern California, Los Angeles,
California 90089-0484, USA}
\date{\today }
\begin{abstract}
Infrared  Hall measurements in the pseudogap phase of the high-$T_c$ cuprates are
addressed within the framework of the ordered $d$-density wave  state. The
zero-temperature Hall frequency $\omega_H$ is computed
as a function of the hole-doping $x$. Our results are consistent with recent experiments in absolute units. We also discuss the signature of the quantum critical point
in the Hall frequency at a critical doping inside the superconducting dome, which can be tested in future experiments.
\end{abstract}
\maketitle

\section{Introduction}

An ordered state  known as the $d$-density wave 
(DDW) has been proposed as the origin of the pseudogap phase of the cuprates.\cite{ck,clmn} 
A variety of experiments have been explored from this perspective.
These include the superfluid density and the resonance peak in neutron scattering,\cite{Sumanta}
the Hall number,\cite{nHall} angle resolved photoemission spectroscopy (ARPES),\cite{ARPES}
the specific heat,\cite{hykee1,chakravarty} the quasiparticle charge,\cite{hykee2} and the direct
signature of DDW in polarized neutron scattering.\cite{neutron1,neutron2,neutron3} In addition, it has been
explored how the notion of this competing order, when combined with interlayer tunneling, and the
doping imbalance of the multilayered cuprates, can result in the striking systematics of the
layer dependence of the superconducting transition temperature $T_{c}$.\cite{ILT}
In all cases, the theory is consistent with the existing observations. 

In this paper, we will address the zero-temperature infrared (IR) Hall angle $\Theta_H$
as a function of the hole-doping $x$, because we are encouraged by the recent measurements
of Rigal {\it et al.}\cite{Rigal} There are two specific reasons: (1) The DDW state predicts
hole pockets as Fermi surfaces in the underdoped cuprates, which should have
important experimental consequences. ARPES experiments
can only detect half of each of these pockets,\cite{ARPES} which therefore appear as Fermi arcs.\cite{foot1}
Thus an important prediction of our theory remains untested, except through its indirect signature in the
doping dependence of the superfluid density. A measurement of $\Theta_H$ can, in principle,
clarify this issue, and we believe that it has.\cite{Rigal} (2) The DDW theory also predicts a
quantum critical point at a doping $x=x_{c}$ within the superconducting dome and
it has been argued that this should be visible in the Hall number, $n_H$,\cite{nHall} if superconductivity
is destroyed by applying a magnetic field. There is some experimental evidence
of this effect.\cite{Boebinger} The difficulty with this experiment is that it needs to
performed in a field as high as 60 T, which is experimentally quite demanding.
We believe that a measurement of ${\Theta_H}(\omega)$ at high frequencies
in the pseudogap state above $T_c$ should have a similar behavior at $x_c$ as
$n_H$  does. We expect that the high frequency behavior at $T>T_c$ will be
similar to the $T=0$ behavior with superconductivity destroyed by a magnetic
field if both experiments probe the same underlying state -- which we believe
is the DDW state -- which causes the pseudogap and coexists with superconductivity
in the underdoped superconducting state.

\section{Mean field formalism of the DDW state}
Given that the DDW state is a broken symmetry state with a local order parameter,
it should be describable by a mean field Hartree-Fock theory and  its consequent
elementary excitations. This is precisely the approach we shall assume in the present paper.  
The mean field Hamiltonian for the DDW state is
\begin{equation}
H = \sum_{{\bf k},\alpha}
[\left(\epsilon_{\bf k}-\mu\right)c^\dagger_{{\bf k} \alpha}c_{{\bf k}
\alpha} +
(iW_{\bf k} c^\dagger_{{\bf k} \alpha}c_{{\bf k+Q}\alpha}+\text{h.
c.})], 
\end{equation}
where $c_{{\bf k}}$ is the annihilation operator for an electron
of spin $\alpha$ in the $z$-direction and momentum ${\bf k}$, $\mu$ is the chemical potential,
and the vector ${\bf Q} = ({\pi}, {\pi})$.  The lattice spacing will be set to unity. 
We ignore the residual interactions between quasiparticles;
the principal effect of electron-electron interactions is to produce non-zero
$W_{\bf k}$.

The single particle spectrum on the square lattice with nearest-neighbor
hopping $t$ and next-neighbor hopping $t'$ is
\begin{equation}
\epsilon_{\bf k}= - 2t(\cos k_{x}+\cos k_{y})+4t'\cos k_{x} \cos k_{y} .
\end{equation}
The $d$-wave order parameter of the DDW state is 
\begin{equation}
W_{\bf k} = \frac{W_{0}(x)}{2} (\cos k_{x}-\cos k_{y}),
\end{equation}
where the amplitude $W_{0}(x)$ is a function of doping.

We can express the Hamiltonian in terms of  a two-component
quasiparticle operator:
$\Psi_{{\bf k},\alpha}^\dagger~=~(c_{{\bf k}\alpha}^\dagger, -i c_{{\bf
k+Q}\alpha}^\dagger)$,
and then diagonalize this
$2\times 2$ Hamiltonian to get
\begin{eqnarray}
H = \sum_{{\bf k},\alpha} \chi_{{\bf k},\alpha}^\dagger
\left(\begin{array}{cc}[E_{+}({\bf k})-\mu] & 0 \\
0 & [E_{-}({\bf k})-\mu]
\end{array} \right) \chi_{{\bf k},\alpha} .
\label{eq:twobytwo}
\end{eqnarray}
The two-component quasiparticle
operator $\chi_{{\bf k}\alpha}$ is  unitarily related to $\Psi_{{\bf k}\alpha}$,
and the sum is over the reduced Brilloun zone (RBZ). 
$E_{\pm}({\bf k}) = \epsilon_{2\bf k} \pm
\sqrt{\epsilon_{1\bf k}^{2} +W_{\bf k}^{2}}$ are the two bands
of the ordered DDW state, with
$\epsilon_{1\bf k} = -2t(\cos k_{x}+\cos k_{y})$ and
$\epsilon_{2\bf k} = 4t'\cos k_{x}\cos k_{y}$.

\section{Calculation of the infrared Hall angle}

For a system of DDW quasiparticles  in the presence of a magnetic field $\bf H$ in the $z$-direction,
 and an electric field $\bf E$ in the $x-y$ plane, $\Theta_H$ is the angle between
 $\bf E$ and the current ${\bf j}$: $ \tan \Theta_H=E_y/E_x=\sigma_{xy}/\sigma_{xx}$.
We will compute the necessary conductivities, $\sigma_{xy}$ and $\sigma_{xx}$
in the framework of Boltzmann theory\cite{Ziman} applied to the DDW mean-field hamiltonian.
Since we consider a non-interacting model, this semiclassical approach easily
generalizes to finite frequencies as well. A number of comments regarding the validity of our Boltzmann approach are in order.
\begin{enumerate}
\item In a normal metal,  it is well known (see Ref.~\onlinecite{Peierls}) that the external frequency $\omega$ and wavevector $q$ must satisfy $\omega\ll \mu$ and $q\ll k_{F}$, where $k_{F}$ is the Fermi wavevector. Although we must have $k_{F}l\gg1$ for localization effects to be neglected ($l$ is the mean free path), there are no further restrictions on the product $\omega\tau$, where $\tau$ is the lifetime due to impurity scattering.
\item In a superconductor, the same conditions apply at high frequencies, unless we want to capture interesting order parameter disequilibrium effects, such as charge imbalance etc., whence we must satisfy $\omega \ll \Delta$, where $\Delta$ is the superconducting gap.\cite{Tinkham} 
\item For a particle-hole condensate, such as DDW, the condition for the  validity of the Boltzmann equation should be the same as in a normal metal.  The diagonalization in
Eq.~(\ref{eq:twobytwo}) does not mix particles and holes and, therefore,
we can apply the Boltzmann formalism to DDW quasiparticles,
which have relatively simple, particle-number conserving scattering terms.
\item We assume that DDW quasiparticles have only
one scattering time, though it may vary along the Fermi surface.\cite{stime,Abrahams}
This assumption is clearly supported by experiments, at least in the pseudogap regime of
YBCO$_{y}$ for $y=6.45-6.75$.\cite{Ando1} It surprisingly  appears to be true for
even very lightly doped sample of $y=6.30$.\cite{Ando2}
The alternate view that for each $\bf k$, there are two scattering
times\cite{Anderson} $\tau_{H}\sim T^{-2}$ and $\tau_{tr}\sim T^{-1}$ appears to be untenable
in this regime. (Above the DDW ordering temperature, the situation may, of course,
be more complicated.)
\item Further complications from interband transitions will be neglected, because, to a first approximation, the effect of these high frequency processes will be simply to renormalize the effective single band parameters.
\end{enumerate}

The longitudinal and Hall conductivities, $\sigma_{xx}(\omega)$ and $\sigma_{xy}(\omega)$,
are: 
\begin{widetext} 
\begin{eqnarray}
\nonumber \sigma_{xy}(\omega) &=& \frac{2e^{3}H}{\hbar^{4}c}
\int_{\text{RBZ}}\frac{d^{2}k}{(2\pi)^{2}}\left(\frac{\tau_{\bf k}}{1-i\omega\tau_{\bf k}}\right)^{2}
\frac{\partial E_{+}(k)}{\partial k_{x}}\Big[\frac{\partial E_{+}(k)}{\partial k_{y}}
\frac{\partial^{2}E_{+}(k)}{\partial k_{x} \partial k_{y}}
  -\frac{\partial E_{+}(k)}{\partial
k_{x}}\frac{\partial^{2}E_{+}(k)}{\partial k_{y}^{2}}\Big]
\delta(E_{+}(k)-\mu)
+ (E_{+} \rightarrow E_{-})\\
 &=& \frac{2e^{3}H}{(2\pi)^{2}\hbar^{4}c}I_{xy}[\omega,\tau_{\bf k}],
\label{eq:xy} \\
\sigma_{xx}(\omega) &=& \frac{2e^{2}}{\hbar^{2}}\int_{\text{RBZ}}\frac{d^{2}k}{(2\pi)^{2}}
\left(\frac{\tau_{\bf k}}{1-i\omega\tau_{\bf k}}\right)\left(\frac{\partial E_{+}(k)}{\partial k_{x}}\right)^{2}
\delta(E_{+}(k)-\mu) +(E_{+} \rightarrow E_{-})
=  \frac{2e^{2}}{(2\pi)^{2}\hbar^{2}} I_{xx}[\omega,\tau_{\bf k}] .
\label{eq:xx}
\end{eqnarray}
\end{widetext}
Here, we have defined
$I_{xy}[\omega,\tau_{\bf k}]$ and $I_{xx}[\omega,\tau_{\bf k}]$ for later reference.
In the equation for $ \sigma_{xy}(\omega)$, we have made the
approximation $\nabla_{\bf k}\ln \tau_{\bf k}\approx 0$, which is very reasonable
so long as $\tau_{\bf k}$ is large and varies smoothly.

Thus, the finite frequency Hall angle is given by,
\begin{equation}
 \tan \Theta_H(\omega)=\frac{\sigma_{xy}(\omega)}{\sigma_{xx}(\omega)}
=\frac{eH}{\hbar^{2}c}\frac{I_{xy}[\omega,\tau_{\bf k}]}{I_{xx}[\omega,\tau_{\bf k}]}.
\label{eq:tanw}
\end{equation}
At finite frequency $\omega$, $\tan \Theta_H (\omega)$ becomes complex.
In the limit that $\omega\tau_{\bf k} \gg 1$, the imaginary part can be determined without the complications\cite{Abrahams} of the unknown anisotropic $\tau_{\bf k}$. Thus,
\begin{equation}
 \text{Im} [\cot\Theta_H(\omega)]=-\frac{\omega}{\omega_H},
\label{eq:cot}
\end{equation}
where $\omega_H$, the Hall frequency, is defined as 
\begin{equation}  
\omega_H=\frac{eH}{\hbar^{2}c}\frac{I_{xy}^{0}}{I_{xx}^{0}},
\label{eq:wh}
\end{equation}
where $I_{xy}^{0}$ and $I_{xx}^{0}$ are the same as the integrals $I_{xy}[\omega,\tau_{\bf k}]$ and $I_{xx}[\omega,\tau_{\bf k}]$, except that the factor $\left(\tau_{\bf k}/[1-i\omega\tau_{\bf k}]\right)$
is replaced by unity. The imaginary part of $\cot\Theta_H(\omega)$ can therefore be determined in a largely
model-independent manner -- in this limit, it is essentially a measure of Fermi
surface geometry \cite{Beaulac} --
while the real part of $\cot \Theta_H(\omega)$ involves
the unknown parameter $\tau_{\bf k}$ which can depend on many details. 

If we had replaced $\tau_{\bf k }$ by its $\bf k$ average over the Fermi surface, $\tau$, the expression for $ \text{Im}[\cot\Theta_H(\omega)]$ would have been exactly the same as in Eq.~(\ref{eq:cot}), regardless of the magnitude of the product $\omega\tau$,\cite{foot2} but now there would have been a dissipative real part containing $\tau$, that is, within this approximation:
\begin{equation}
\cot\Theta_H(\omega)=\frac{1}{\omega_{H}\tau}-i\frac{\omega}{\omega_{H}}
\end{equation}

\section{Results}

With Eq.~ (\ref{eq:wh}) in hand, we can now calculate $\omega_H$ as a function of $x$ in the
DDW state. We choose a representative set of values for the needed parameters. In keeping
with our analysis for the related quantity, $n_{Hall}$, \cite{nHall} we choose
$t=0.3$ eV, $t'/t = 0.3$. For a comparison with experimental data, it is necessary to
choose an appropriate relation between the chemical potential $\mu$ and the doping $x$.
Physically, this relation can be exceedingly complex in the underdoped regime, where a
plethora of competing charge and spin ordered states can intervene as $x\to 0$.
We, therefore, do not discuss the behavior in this heavily underdoped regime,
although in the past we have attempted to describe this regime by arguing that the
chemical potential is perhaps  pinned to zero.\cite{Sumanta} Between the
overdoped ($x \gtrsim 0.2$) and the moderately underdoped regime ($x \sim 0.07$),
we make the simplest possible assumption that $\mu$ is a smooth function of $x$.
The actual function is not very significant, but to be concrete, we choose the relation
implied by the band structure. To discuss the nonanalyticity close to $x_{c}$,
we neglect the $x$ dependence of the chemical potential for simplicity,
as the doping dependence of $W_{0}(x)$ is much more important.
For illustrative purposes, we take
\begin{equation}
W_{0}(x)=0.03[(1-x/{x_c})+(1-x/{x_c})^{1/2}] \; \text{eV}
\label{eq:W0}
\end{equation}
with ${x_c}=0.2$. This form gives mean-field-like
nonanalyticity at $x_c$. (The exponent $1/2$ can be replaced if, for instance,
$3D$ Ising behavior is preferred.) It is also a reasonable representation of the  form suggested in Ref.~\onlinecite{Loram}. We believe, however,  that the final result is not
strongly dependent on this particular detailed form of Eq. ~(\ref{eq:W0}).

\begin{figure}[htb]
\centerline{\includegraphics[scale=0.4]{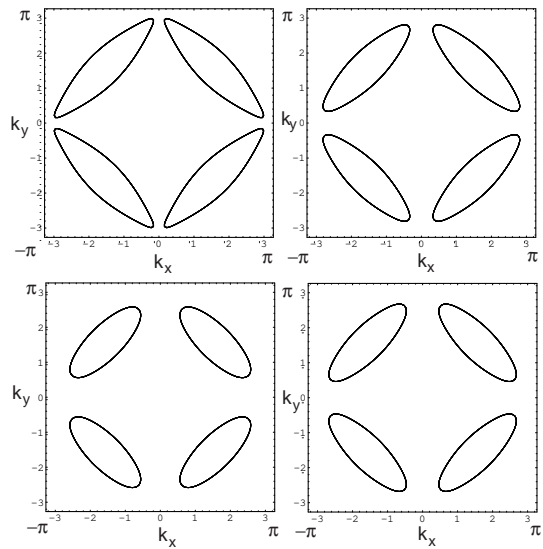}}
\caption{Contour plot of the hole pockets at the fixed value of the chemical
potential $\mu=- 0.36\; \text{eV}$ as the DDW gap is varied (implicitly as a function of doping $x$).
Proceeding clockwise from the left panel on the top, $W_{0}(x)=0.01, 0.05, 0.1, 0.15\; \text{eV}$; these numbers are for illustrative
purposes only, and the $\mu$ was chosen so as to focus on the lower band only.}
\label{fig:cplot}
\end{figure}

Before doing an
explicit calculation, it is revealing to plot the hole pockets as the amplitude of the
DDW gap $W_{0}(x)$ is increased for $\mu$ varying smoothly as a function of
doping $x$, in particular we show the results for a {\em fixed} value of 
$\mu$ in Fig.~\ref{fig:cplot}. One can see that the
hole pockets become less elliptical as $W_{0}$ is increased at constant $\mu$.
So, even though $\mu$ is kept constant, with increasing $W_{0}$ below $x_c$,
the curvature of the holepockets increases where the Fermi velocity is largest, \cite{Beaulac}
consequently  $I_{xy}$ increases, and the
perimeter decreases, so that  $I_{xx}$ decreases. \cite{Beaulac}
The net result is an increase of $\omega_{H}$.
This is a robust explanation of the increase of $\omega_{H}$ as the system is underdoped in
agreement with Rigal {\em et al.}\cite{Rigal}

\begin{figure}[htb]
\includegraphics[scale=0.5]{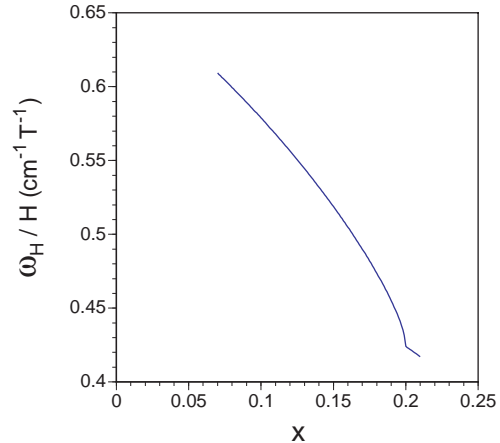}
\caption{ The Hall frequency ${\omega}_H$ as a function of $x$. The choices of the DDW gap amplitude and the chemical potential are described in the text.}
\label{fig:omega}
\end{figure}
 
To be quantitative, we explicitly calculate $\omega_{H}$ using Eq.~(\ref{eq:W0}) and the band
structure parameters given above. Concomitantly, as mentioned above, $\mu$ was determined from the band structure.
The results are shown in Fig.~\ref{fig:omega}.
The results are clearly consistent with the experiment of Rigal {\em et al.}\cite{Rigal} The enhancement, as the system is underdoped, is significant, even though its actual magnitude is perhaps a factor of two smaller. Moreover, the absolute magnitudes are well captured. Beyond this, it is difficult to compare in detail. The experiment was performed on thin films of YBCO for which we do not have  the precise knowledge of the doping levels, nor do we have a good criterion to relate the doping with the chemical potential. To complicate matters further, the chain contributions in YBCO are not included in our calculation, and these contributions were not subtracted in their experimental results. The parameters used here are generic; it is possible to improve the agreement with the experiment by adjusting them, but we do not find this to be a very meaningful exercise.

\begin{figure}[htb]
\includegraphics[scale=0.5]{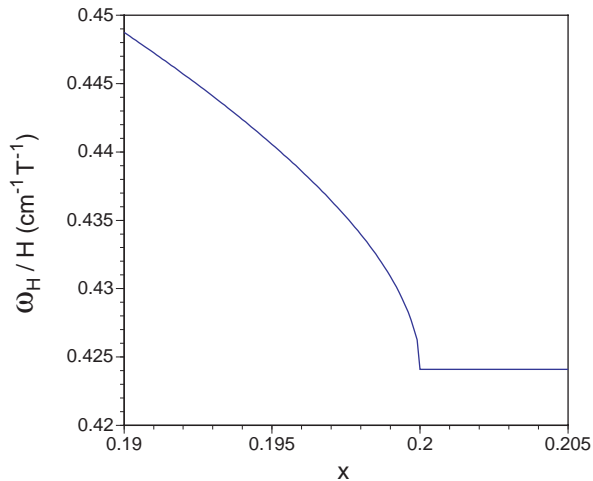}
\caption{The Hall frequency ${\omega}_H$ close to the quantum critical point $x_{c}=0.2$,
with $\mu$ set to a constant, equal to $-0.3$ eV.}
\label{fig:omega2}
\end{figure}

We then explore more closely the signature of the quantum critical point in the infrared Hall angle,
which is already evident in Fig.~\ref{fig:omega}. To demonstrate the robustness of the  quantum critical point, 
we  set $\mu$ to be a constant, equal to $-t = -0.3 \; \text{eV}$ 
and use  the mean field ansatz $W_0(x)=W_0(1-x/x_{c})^{1/2}$.
For a representative value, we take $W_0=0.03 \; \text{eV}$ and ${x_c}=0.2$. These are identical to our previous parametrizations.\cite{nHall}
The results are shown in Fig.~\ref{fig:omega2}.

By examining  the integrals $I_{xy}$ and $I_{xx}$ in more detail, we find that, close to the critical doping, $\omega_H(x)=C_1 + {C_2}{W_0}(x_c-x)^{1/2}$, where $C_{1}$ and $C_{2}$ are constants,
and the slope diverges at the transition. (At finite temperature, this will be rounded.)
We emphasize that although the hot spots determine the critical singularity close to $x_{c}$,\cite{nHall} the increase of $\omega_{H}$ is determined by the evolution of the hole pockets and the gapping of the Fermi surface as the doping is decreased.

\section{Conclusion}

The Hall angle measurements
of Rigal, {\it et al.} \cite{Rigal} are strong
evidence for the existence of hole pockets
in the underdoped cuprates. We believe that
DDW order is the simplest
explanation which is also consistent
with the absence of hole pockets in ARPES.
Consistency with both of these experiments
(and the Hall number measurement of Ref.
\onlinecite{Boebinger}) is a strong challenge of
other proposals for the pseudogap.
Our analysis opens a number of interesting
directions for future research. Our calculation
could be extended using the Kubo formalism,
which would have a wider region of validity.
A more careful comparison between theory and experiment could
be made with a better model for the chemical potential
at the precise doping levels of the experiment.
Finally, further exploration of the putative quantum critical
point at $x=x_c$ at which DDW order vanishes
could more firmly establish its existence
and its properties.

 \acknowledgements
We thank H. D. Drew for drawing our attention to their experiments and also for useful discussions.
We are also grateful to N. P. Armitage, D. Basov, and S. Kivelson for helpful comments.
C. N. has been supported by the NSF under
Grant No. DMR-9983544 and the A.P. Sloan Foundation.
S. C. has been supported by the NSF under
Grant No. DMR-9971138.  S. T. has been supported in part by the NSF
under Grant No. DMR-9971138, and in part by DMR-9971138.
J. F. was supported by the funds from the David Saxon chair at UCLA.


\begin{thebibliography}{99}
\bibitem{foot1} One might be concerned that our model is in contradiction
with claims that the nodal fermi velocity is independent of doping in copper oxide superconductors [X. J. Zhou {\em et al.}, Nature {\bf 423}, 398 (2003)]. An incorrect, naive
argument can be constructed to show that the Fermi velocity will
vary as $\sqrt{x}$ with doping $x$ for any model with hole
pockets -- thereby supposedly ruling out such models. If we assume that
the hole pockets are circular, then it might be argued that the $k_{F}$ associated with the hole pocket will lead to the relation $k_{F}^{2}\propto x$; hence, the nodal Fermi velocity should be given by $v_{F}\propto \sqrt{x}$, assuming that the effective mass is unchanged. That this is simply an incorrect argument can be seen from Ref.~\onlinecite{ARPES}. It is trivially seen that, along the nodal direction, $k_{F}$ seen in ARPES is the same as that predicted by the band structure because of the special properties of the coherence factors.\cite{ARPES} This is approximately true even away from this direction, as was verified by explicit calculations. In any case, we also do not find the experimental evidence of the doping independence of $v_F$ convincing: a collage of measurements of chemically diverse materials, in each of which $v_{F}$ has significant, but conflicting doping dependence, was used to argue that statistically there is no overall doping dependence. Finally, in the same experimental paper mentioned in this footnote, it was also argued that the `high-frequency Fermi velocity' diverges as $\frac{1}{x}$, as the doping $x \to 0$. This is not a very meaningful statement, as at higher energies there are no quasiparticles to speak of in the ARPES spectra.
\end{thebibliography}
\end{document}